\documentclass[journal]{IEEEtran}
\usepackage{amsmath,amssymb,comment,amsthm,comment}
\usepackage{caption}

\usepackage{cite}
\usepackage{multirow}
\usepackage{graphicx}
\usepackage{hyperref,tikz}
\theoremstyle{boldthm}
\newtheorem{theorem}{Theorem}
\usepackage{pifont} \usepackage{booktabs} \usepackage{colortbl} \usepackage{xcolor}
 \definecolor{headerblue}{HTML}{1a3a5c}
 \definecolor{groupblue}{HTML}{d6e4f0}
 \definecolor{checkgreen}{HTML}{1a7a3c}
\definecolor{crossred}{HTML}{c0392b}
 \definecolor{rowalt}{HTML}{f5f9fc}

\newcommand{\cmark}{\textcolor{checkgreen}{\textbf{\ding{51}}}}
\newcommand{\xmark}{\textcolor{crossred}{\textbf{\ding{55}}}}

\title{From Passive Mirrors to Active Agents: Holonic Digital Twins for Physical AI over Networks}
\author{Christo Kurisummoottil Thomas, \emph{Senior Member, IEEE,} Omar Hashash, \emph{Member, IEEE,}\\ and
		Walid Saad,  \emph{Fellow, IEEE.} \vspace{-6mm}\thanks{Christo Kurisummoottil Thomas is with Department of Electrical and Computer Engineering, Worcester Polytechnic Institute, Worcester, MA, USA, Walid Saad and Omar Hashash are with the  Bradley Department of Electrical and Computer Engineering,
Virginia Tech, Alexandria, VA, USA. Emails:\{cthomas2@wpi.edu, omarnh@vt.edu and walids@vt.edu\}}\thanks{The work of Omar Hashash and Walid Saad were supported by the U.S. National Science Foundation under Grant
 CNS-2210254.} \vspace{-6mm}}

\begin{document}

\maketitle

\begin{abstract}
Artificial intelligence (AI) revolutionized multiple sectors ranging from healthcare to entertainment. Remarkably, despite this progress, today's AI tools, including deep learning  and generative AI (e.g., large language models), still fail when embedded into physical systems, such as robots or vehicles that, operate under the physical laws of the real world. 
This limitation stems from the inability of physical AI systems to maintain reliable world models for long-horizon planning under uncertainty and to generalize to unseen scenarios. In this context, wireless networks, through their pervasive sensing and communication infrastructure, can act as orchestrators of physical intelligence. However, current wireless architectures optimize for
throughput, latency, and reliability and cannot
support real-time physical AI coordination where
agents must maintain a shared spatiotemporal context. To overcome such challenges, in this paper, a network of holonic digital twins (HDT-Nets) framework is proposed to deliver  the real-time physical AI inference requirements
through \emph{holonic agents} that actively reason about
their environment rather than passively mirror
physical assets. Each HDT is realized as a hierarchical structure
spanning the physical agent and the network edge,
reasoning autonomously at the local level while
cooperating with neighboring HDTs to form larger
collectively intelligent units at the network level. In HDT-Net, causal Markov blankets (MBs) spanning
sensing, communication, and control domains
determine which agents must coordinate and enable
counterfactual reasoning over multi-domain
interventions. Active inference, operating within
these boundaries, unifies perception, action, and
learning by minimizing expected free energy,
simultaneously deciding what beliefs to transmit
based on their cognitive value to the receiver.
Category theory guarantees that these transmitted
beliefs preserve semantic structure when crossing
between heterogeneous agents with incompatible
representations. Finally, integrated information theory provides a quantifiable metric to capture  when collective intelligence exceeds independent operation and how network intelligence evolves through coordinated learning and  information exchange.
\end{abstract}

\vspace{-4mm}\section{Introduction}
\label{sec:introduction}\vspace{-2mm}

% 6G components, State of the art on digital twins, digital twins as simulators, LLMs or diffusion models, DTs with world models, transition to physical AI, new revenue stream by coping chatgpt tokenization to DT with reasoning agents. Real definition of twin in 6G? How do we implement or realize this DTs with reaosning? Intersection of physical AI with DTs.
%Artificial intelligence (AI) revolutionized multiple sectors ranging from healthcare to entertainment. Remarkably, despite this progress, today's AI tools, including deep learning and generative AI (e.g., large language models) \cite{BariahEtAl2024GenAI4Telecom}, still fail when embedded into physical systems, such as robots, drones, or vehicles, that operate under the physical laws of the real world, as evidenced by recent high-profile incidents involving Waymo, GM's Cruise, and Tesla Autopilot.  Indeed, the success
%of physical AI systems is contingent upon addressing intertwined
%challenges: (a) Limited ability of existing AI frameworks to handle unseen and out-of-domain scenarios when deployed at resource-constrained edge nodes, (b) Lack of first-principle solutions enabling edge AI agents to understand physics that generates the data and reason about rather than merely pattern-match from training data, and (c) Need for pervasive
%connectivity to support physical AI tasks, such as inference
%and communications, at scale. %Need for principled networking frameworks coordinating heterogeneous edge agents whose beliefs about the world must fuse coherently despite incompatible sensor modalities and AI models.

Physical AI represents systems that do not just process information in the digital realm but they must  perceive, reason, and act under the laws of the physical world \cite{saad2025artificial}. Unlike current generations of AI that operate on curated datasets and well-defined input-output spaces, physical AI systems must be able to perform spatial and temporal reasoning, act under dynamic uncertainty, coordinate with other physical agents, and continuously adapt to a dynamic world in real-time. 
Such requirements become indispensable because the
physical world does not tolerate failures, whether
caused by a single agent's perception errors or by
emergent conflicts when multiple agents interact in
a shared environment.
%Examples of these AI failures can include GM Cruise's suspension of operations following a pedestrian collision, Waymo vehicles exhibiting unexpected behavior near traffic control devices, and multiple fatalities linked to Tesla Autopilot's failure to handle out-of-distribution scenarios \cite{nhtsa2024crashreport}.

The root causes of such failures stem from several limitations. First, physical AI systems often  \emph{do not have a  reliable model of themselves or their environment} as physical actors. They cannot predict how their own actions will play out in the real world, and therefore cannot always plan for failure, adapt to changing conditions, or recover when reality diverges from expectation. Second, today's AI models are \emph{data-driven models that learn statistical correlations} from training data but have limited ability to reason about the effect of change caused by their own actions and change caused by exogenous disturbances \cite{thomas2023causal}. %Second, AI systems \emph{lack physics-grounded world models}: they do not operate from first principles governing how physical entities move, interact, or constrain each other in space and time. For example, a vision model can recognize an object but cannot predict its ballistic trajectory when dropped; a language model can describe friction but cannot reason about whether a robot's gripper force will secure or crush a fragile payload. This absence of physics-based reasoning means AI cannot reliably predict the physical consequences of its actions or recover when reality violates learned statistical patterns. 
Third, physical tasks require \emph{long-horizon planning under deep uncertainty} stemming from physical laws. Yet current AI architectures cannot maintain accurate beliefs about the world, grounded in its governing laws, over the extended time horizons that physical tasks require. Fourth, models trained in simulations encounter \emph{distribution and domain shift} upon deployment in the real world governed by physical laws that give rise to unforeseen scenarios.
 %The root cause extends beyond algorithmic limitations: physical AI systems face a fundamental infrastructure mismatch.  
No single physical AI agent can sense its entire environment,  sustain long-horizon inference under real-time constraints, or recover from failure in isolation. %The above limitations motivates a fundamental rethinking of the role of wireless cellular systems such as 6G. 
Herein, due to their pervasiveness and access to heterogeneous sensing infrastructure, wireless  cellular systems such as 6G  have a unique opportunity to act, not only as communication providers, but rather as \emph{orchestrators of physical intelligence}. Particularly, due to their pervasive deployment
across the same physical environments in which agents
operate, 6G systems can serve as a cognitive medium
that enables collective inference at scale, allowing
physically distributed agents to pool observations,
coordinate actions, and reason jointly about a shared
world that no single agent can fully observe. However, current wireless networks are typically designed as passive data transport layers, and, thus, they still cannot support the \emph{real-time inference and collective reasoning} that is required to operate and coordinate physical AI at scale. This is because the \emph{real-time operations} required of closed-loop physical action are categorically different from those of streaming or cloud inference. Any inference update communicated over the air but \emph{not synchronized perfectly with the real world dynamics}  could result in hazardous consequences on the \emph{safe operation of physical AI} systems \cite{popovski2024time}. This requires the wireless  network to provide a shared spatiotemporal context among agents, ensuring that distributed belief updates, sensing, and control actions remain coherent across space and time. 

The above challenge has catalyzed a fundamental architectural shift. As envisioned in our prior work \cite{saad2025artificial}, network infrastructure must transition from  \emph{pure data pipes into distributed inference and reasoning engines} by deploying AI compute directly at the network edge.  The core challenge here is enabling networked agents to reason collectively, coordinate beliefs, and act under physical constraints. 
 This, in turn, requires the network to actively participate in the \emph{perception-decision-action} loop by providing a shared spatiotemporal context that ensures agents perceive the same physical reality at the same moment, make decisions grounded in synchronized world models, and commit to coordinated actions within the tight timing windows that safe physical operation demands. This results in following challenges in future network design. First, current networks have not addressed how transmission decisions should be governed by the \emph{cognitive value} a packet carries to a receiving agent operating under uncertainty in the physical world.  Unlike conventional notions on value of information \cite{Howard1966InformationValue}, cognitive value measures how much a transmitted belief improves the receiving agent's world model about causally relevant hidden states, thereby enabling accurate  sensing, timely communication, and effective  control decisions in the physical environment. Cognitive value is critical for physical AI systems since an agent operating on an inaccurate or  outdated world model risks irreversible physical harm, making the cognitive value of each transmission a safety-critical quantity rather than a traditional quality-of-service measure. Second, closed-loop physical action requires that inference updates are \emph{perfectly synchronized with real-world dynamics} since any misalignment between communicated beliefs and the physical state of the world can result in hazardous consequences for safe operation~\cite{popovski2024time}.
 Third, physical AI agents with different sensor modalities and AI architectures encode the world in \emph{incompatible representational spaces}. However, existing network frameworks provide no mechanism for aligning, or composing these heterogeneous representations in a way that preserves their semantic structure for collective inference. Finally, and most fundamentally, the network \emph{lacks theory of mind} \cite{Wang2026MetaMind} and \cite{Wang2026DMWM} capabilities that can allow it to predict how physical AI agents will interpret and act upon transmitted beliefs, and cannot adapt their computing, sensing and communication resources to agents' evolving cognitive states and coordination requirements.

Overcoming these challenges requires the network to evolve into a \emph{cognitive infrastructure} that (a) prioritizes transmissions based on their cognitive value to downstream agents, (b) maintains tight spatiotemporal synchronization between communicated beliefs and physical dynamics, (c) aligns heterogeneous representations into a shared semantic space for collective inference, and (d) models and adapts to the evolving beliefs and actions of agents to enable coordinated decision-making in real time. The transition to a cognitive infrastructure requires a computational abstraction in which representations of the physical world, agent states, and their causal interactions can be \emph{continuously inferred, synchronized, and acted} upon across the network. \emph{Digital twins (DTs) \cite{khan2022digital} naturally provide such an abstraction.} By maintaining virtual representations of physical agents and their environment, DTs allow physical AI systems to reason about future outcomes, align heterogeneous observations, and support coordinated decision-making. At their core, these capabilities are enabled by an underlying world model that defines how the twin represents the world and captures its underlying dynamics. \emph{A world model \cite{LeCun2022Path} is essentially the computational substrate in which the DT lives and can be defined as a structured representation of physics, geometry, causality, and uncertainty that enables agents to mentally simulate the consequences of actions and reason about what would happen under interventions rather than merely predicting what has been observed.} However, existing DT implementations \cite{Manalastas2024Simulators,Masaracchia2023OpenRAN,Plageras2022DigitalTwinsIIoT,NguyenKha2026DigitalTwinDSS,khan2022digital, li2025generative, haider2025llmdt,hoydis2024learning,alikhani2024lwm} treat twins primarily as passive simulators for prediction or offline optimization. Such a design for DTs is not suitable for physical AI systems that must maintain accurate beliefs about a rapidly changing physical world, act under deep uncertainty about hidden states, and coordinate with neighboring agents within the tight spatiotemporal constraints that safe physical operation demands. %In this regard, DTs must evolve into active reasoning entities that maintain causal world models, quantify and actively reduce their own uncertainty using active sensing, and support counterfactual inference about the consequences of actions before they are executed in the physical world. 
Realizing the vision of cognitive DTs that actively
perceive, infer, and act within a shared physical world
naturally leads to the \emph{emergence of collective
intelligence}, where individual HDTs compose into
progressively larger cognitive units whose integrated
world models capture world structure that no single
agent can represent alone. However, achieving this
collective intelligence means that the
network must actively direct sensing resources toward
observations that most reduce collective uncertainty,
maintain safety-critical control loops where
millisecond-scale delays cause physical catastrophes,
and support belief exchange where agents transmit
inferred beliefs about hidden states rather than raw
sensor measurements. These capabilities must hold even
under damaged infrastructure, environmental
non-stationarity, and mission-critical time constraints.
Meeting these requirements motivates a fundamentally
new network architecture in which every physical AI
agent is paired with a cognitive twin that reasons,
senses, and coordinates on its behalf.
\vspace{-4mm}\subsection{Key Contribution: A Vision of Holonic Digital Twins}
\begin{figure*}[t]
\centering
\includegraphics[width=1.6\columnwidth]{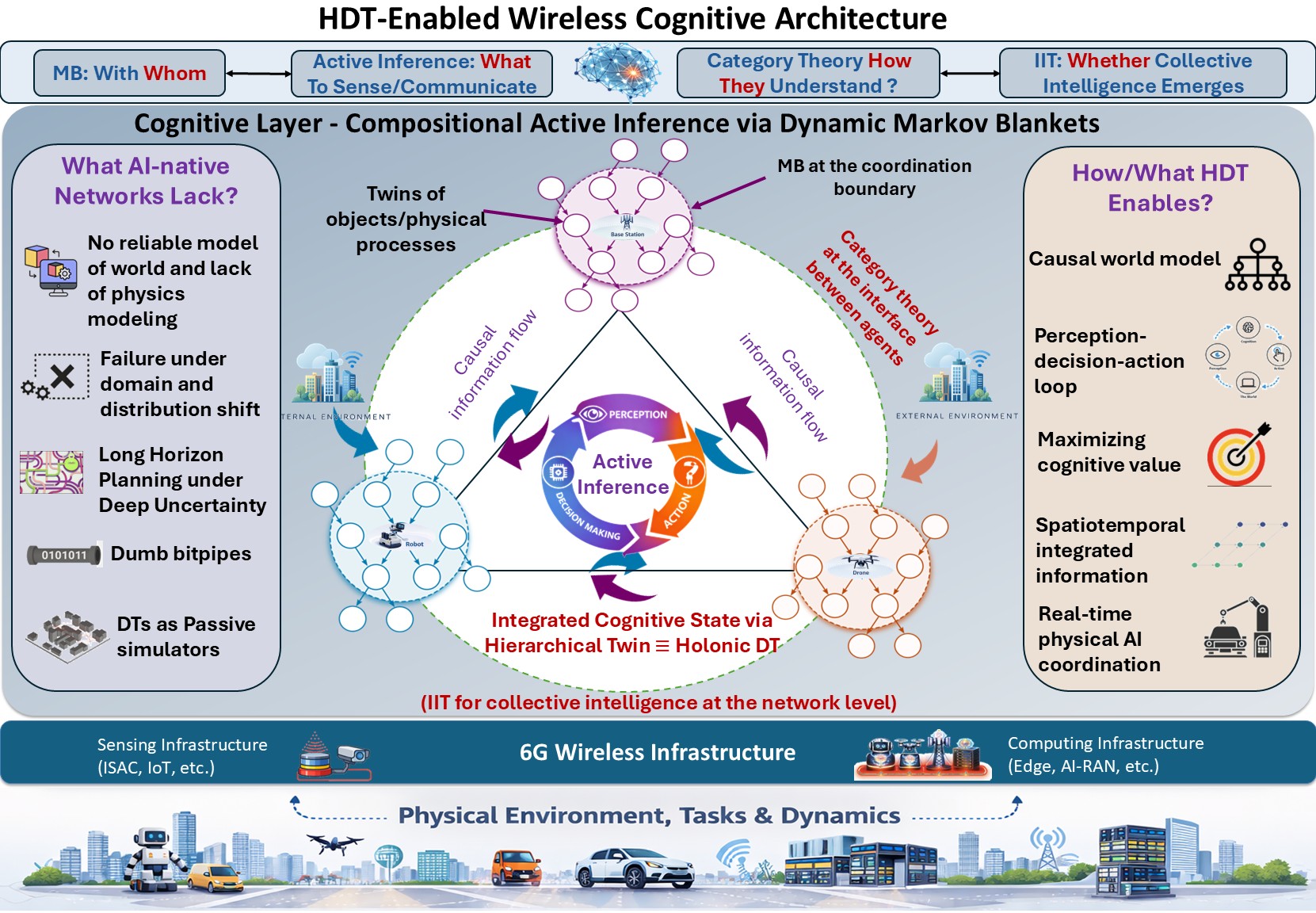}
\caption{\centering\small Overview of Proposed HDT-Net.}
\label{BlockDiagram}
\vspace{-5mm}\end{figure*}

\vspace{-2mm}
The main contribution of this paper is to present the transformative concept of a networked holonic digital twins (HDTs), called HDT-Nets, that can address the aforementioned challenges. The term \emph{holonic}, originating from Koestler's
synthesis of the Greek \emph{holos} (whole) and the
suffix \emph{on} (part)~\cite{koestler1968ghost}, captures
the defining structural property of our framework, where each twin is
simultaneously an autonomous cognitive agent that
maintains its own causal world model and reasons
independently about its local environment, and a
cooperative part of progressively larger cognitive
units whose collective intelligence exceeds what any
individual achieves alone. The \emph{6G network orchestrates
this holonic composition} by providing the shared
spatiotemporal context, communication resources, and
edge computation through which individual HDTs
discover coordination partners, exchange beliefs about the world, and
compose into collectively intelligent agents at the network level. We adopt a holonic architecture because physical AI
coordination satisfies all three conditions under
which holonic decomposition has been established as
the only viable structural
principle~\cite{simon1996sciences,valckenaers2008fundamentals}. This includes
bounded onboard computation that limits each agent's
reasoning capacity, dynamic environments that induce domain and distribution shift
faster than any data-driven model can retrain, and demanding safety-critical constraints
where failure to coordinate in real time leads to
irreversible physical harm. As illustrated in
Fig.~\ref{BlockDiagram}, in our envisioned HDT-Net,
every physical AI agent is paired with a holonic
digital twin realized as a hierarchical twin spanning both the physical agent and the network edge, where the agent-side twin maintains a causal world model grounded in local sensor observations, while the network-side twin enables real-time distributed inference through collective belief coordination, multi-agent fusion, and resource orchestration across the 6G infrastructure. %Together, they understand the intent of communication, reason over data using the network infrastructure, and coordinate with neighboring HDTs to resolve collective uncertainty in real time. %Unlike traditional DTs that passively simulate or visualize physical assets, HDTs understand the physics behind the world operations using a causal generative model that enables interventional and counterfactual reasoning over the cause effect variables that describe the physical processes. This causal world model allows the agents to infer hidden world states from partial observations, predict causal consequences of potential actions before execution, and exchange semantic beliefs with other twins to enable collective intelligence that exceeds what isolated agents can achieve. This pairing of physical agents with HDTs fundamentally transforms the network's role by becoming the cognitive medium through which twins develop shared situational awareness via belief exchange. Here, communication itself is guided by theory of mind, meaning that agents decide what to communicate based on reasoning about other agents’ beliefs, knowledge gaps, and likely interpretations.  
To enable these capabilities, we construct HDTs through \emph{four interconnected fundamental pillars that synthesize causal reasoning, compositional structure, adaptive learning, and collective coordination.}  %First, HDTs exhibit \emph{causal understanding}, maintaining generative world models that encode cause and effect structure among distinct network and control variables rather than mere statistical associations. This enables reasoning about counterfactual queries critical for physical action: ``What will causally happen if this robot executes this trajectory in this novel environment?" rather than ``What typically correlates with this sensor pattern in training data?" Second, they provide \emph{principled epistemic awareness}, explicitly quantifying uncertainty over beliefs and recognizing when predictions represent reliable knowledge versus dangerous extrapolation beyond training distributions. This addresses the distribution shift problem by enabling agents to know what they do not know, triggering cautious exploration rather than overconfident action when facing novel physical scenarios. Third, they enable \emph{compositional coordination} with formal guarantees, providing mathematical frameworks for how network should compose beliefs from heterogeneous agents with incompatible representations, when tight integration genuinely improves collective performance, and how collective intelligence evolves as agents learn through physical interaction.

First, real-time distributed inference requires the network to dynamically partition into autonomous reasoning agents based on the causal structure of the underlying cyber-physical system, determining which agents must coordinate, what information must cross agent boundaries, and when coordination is worth its communication cost.
To formalize this, \emph{causal Markov blankets (MBs)} \cite{Pellet2008MarkovBlanketsCausal} provide a principled framework for partitioning cyber-physical networks into autonomous AI agents by defining the
\emph{boundary} of each HDT. In particular, MBs identify the minimal set of variables that must be exchanged for coordination while preserving conditional independence from the rest of the system. These blankets define physical AI agent boundaries by identifying sensory states (observations flowing in), active states (actions flowing out), and internal states (beliefs about the world maintained within). %First, they lack principled mechanisms for determining communication boundaries adaptively as the physical agent environment and tasks change. 
%This is critical, as real-time distributed inference requires identifying the minimal, causally relevant information to exchange among agents. 
Further, MBs allow deciding with whom to coordinate based on causal relevance with respect to physical AI task, when to initiate coordination to meet safety-critical deadlines, and how much information to exchange to balance decision quality against communication latency. Here, the beliefs are computed using a causal generative model that mimics the physics that lead to generation of the observed data.  Further, causality enables counterfactual reasoning via do-calculus that answers ``what would happen if" queries essential for physical AI control. 

Second, scalable physical AI coordination requires that heterogeneous agents with incompatible sensor modalities and world model representations compose their beliefs coherently without sacrificing semantic consistency. This is known as \emph{compositionality}, without which, independently developed agents may each reason correctly in isolation yet produce contradictory joint decisions when coordinating in the shared physical environment. \emph{Category theory} \cite{barr1990category} provides the mathematical foundation for this compositionality and operates at the \emph{interfaces between HDTs}. It provides formal guarantees that semantic structure involving the causal relationships, temporal ordering, and geometric structure is preserved when heterogeneous agents coordinate, regardless of their internal implementation differences. Third, while MBs define agent boundaries and category theory ensures compositional guarantees, we still require a principled mechanism for how each HDT perceives its environment, directs sensing resources, selects communication and control actions, and continuously learns from experience.  \emph{Active inference}, operating \emph{on
the causal generative world model}, provides this unified framework by casting perception, sensing, action selection, and learning as a single optimization objective \cite{friston2017active}. Intuitively, an agent maintains an internal model of the world and selects actions that both achieve its goals and reduce uncertainty about its environment. This is formalized through the minimization of expected free energy, which can be interpreted as a combination of “goal mismatch” (how far outcomes are from desired ones) and “uncertainty” (how unsure the agent is about the world). Consequently, when the agent’s model is confident, minimizing this objective drives goal-directed behavior; when uncertainty is high, it instead drives active sensing and information gathering to improve the model before acting. %Each HDT minimizes expected free energy, which automatically drives it toward goal-directed behavior when its world model is confident, and toward active sensing and information-gathering when uncertainty about the physical environment is high. 
This eliminates the need for separately engineered perception, planning, sensing, and control subsystems, replacing them with a single principled objective grounded in the causal generative model of the physical world. 

Fourth, coordinating multiple active inference agents through wireless networks, however, requires a measure of coordination effectiveness that quantifies whether collective intelligence emerges through sensing and communication or merely consumes resources. Traditional wireless networks optimize physical-layer key performance indicators (KPIs), whereby MAC schedulers and routing protocols cannot distinguish between transmissions that genuinely improve collective intelligence and those whose coordination overhead degrades performance. This leads to either bandwidth waste through excessive belief synchronization impacting physical AI real-time operation or mission failures from insufficient information sharing. Towards this end, the use of  \emph{integrated information theory (IIT)} \cite{tononi2016integrated} can address this challenge  by deriving a  \emph{network-level measure} called spatiotemporal integrated information
$\Phi$
 computed over the
collective state of multiple HDTs. $\Phi$ quantifies how much a network of coordinating agents collectively exceeds the sum of what each agent could achieve independently in terms of resolving uncertainty about the shared physical environment and executing coordinated actions toward mission goals.  We extend IIT with collective intelligence growth dynamics, formalizing how $\Phi$ evolves through continuous learning, physical coupling where agents share constraints or environment, and belief exchanges. Furthermore, we propose that future 6G networks must treat sensing, computation, and communication resources not merely as infrastructure to be optimized for physical layer quality of service (QoS) metrics but as instruments for actively cultivating collective intelligence, allocating them in proportion to their contribution to ${\Phi}$ rather than to immediate task performance. %This combination of four pillars enables HDT-Net to realize collective intelligence via distributed real-time inference for physical AI coordination.

Overall, HDT-Net realizes a network where every physical AI agent is paired with a cognitive twin that lives within a causal world model, actively directs sensing resources toward observations that most reduce uncertainty about hidden world states, and coordinates beliefs with neighboring twins through causally grounded wireless interfaces. The network thereby serves as a \emph{distributed
System~2 reasoning engine}~\cite{kahneman2011thinking}
that elevates physical AI coordination beyond fast
reactive control to deliberative causal inference
over MB boundaries, enabling agents to
reason about hidden states, anticipate the
consequences of actions before committing to them,
and coordinate through shared world models rather
than raw sensory reflexes. This
architecture enables three capabilities essential
for physical AI at scale: a) a \emph{shared spatiotemporal
context} that ensures distributed agents perceive,
decide, and act within a common physical reference
frame; b) \emph{safety-critical coordination} where the
cognitive value of every transmission is evaluated
against its impact on downstream physical actions;
and c) \emph{seamless navigation of the physical world through
pervasive connectivity} that sustains collective
intelligence even as agents move, missions evolve,
and infrastructure degrades.
In summary, it is imperative to design  networks that must evolve from passive data processors to active cognitive systems that reason about their environment, understand their own uncertainty and that of others, coordinate through meaning rather than bits, and modulate integration based on principled measures of collective intelligence. HDT-Net provides the theoretical framework and architectural blueprint that facilitates this transformation.
%\vspace{-2mm}\subsection{Paper Organization}\vspace{-2mm}

The rest of this paper is organized as follows. Section~\ref{sec:limitations} critiques current DT approaches and establishes the case for cognitive agents. Section~\ref{sec:markov_blankets} formalizes causal MBs for network partitioning and compositional reasoning. Section~\ref{sec:active_inference} develops active inference for wireless networks, showing how it enables perception-action loops and introduces IIT for multi-agent coordination.   Section~\ref{sec:conclusion} concludes with open research directions.

\vspace{-4mm}\section{From Passive Mirrors to Active Agents:  \\The Case for Holonic Digital Twins}\vspace{-2mm}
\label{sec:limitations}

\subsection{HDT-Nets over Wireless Networks}

%The DT concept that creates a virtual replica of a physical system for monitoring, prediction, and control optimization has gained significant interests in wireless network research. %Current implementations span radio access network (RAN) optimization \cite{Manalastas2024Simulators}, network planning \cite{Masaracchia2023OpenRAN}, edge computing orchestration \cite{Plageras2022DigitalTwinsIIoT}, and spectrum management \cite{NguyenKha2026DigitalTwinDSS}. 
The holonic DTs in HDT-Net are \emph{twins of
physical AI agents, such as robots, drones, and
vehicles, and the physical environment in which they
operate, not merely twins of the wireless network
itself}. The 6G wireless network facilitates the operation of HDT-Net by serving as the
cognitive medium through which HDTs exchange beliefs and
collectively reason about the shared physical world.
The overall system  comprises physical agents and
network infrastructure, a communication layer that
streams sensor measurements and state updates to
virtual replicas, and a cyber component that
maintains digital representations of the physical
system's state for prediction and optimization.
Using the DT’s simulation capability to evaluate hypothetical scenarios and predict their outcomes, operators can test new physical AI control configurations, simulate fault scenarios, and optimize parameters in a risk-free digital environment before applying changes to the live network. However, despite these
advantages, current DT implementations lack the
holonic property. This is because they cannot reason autonomously
about their own uncertainty, compose with
heterogeneous neighbors into larger cognitive units,
or adapt their coordination boundaries as missions
evolve. This limits traditional DT's ability to support trustworthy and real-time
physical AI operation.

\vspace{-4mm}\subsection{Fundamental Limitations of Current Approaches}
\vspace{-1mm}
\subsubsection{Limitation 1: Passive Observation Without Active Cognition}

Existing DT frameworks for wireless networks~\cite{Manalastas2024Simulators,Masaracchia2023OpenRAN,Plageras2022DigitalTwinsIIoT,NguyenKha2026DigitalTwinDSS,khan2022digital, li2025generative, haider2025llmdt,hoydis2024learning,alikhani2024lwm} focus on synchronization between physical and virtual replicas and emphasizes simulation fidelity for visualization, predictive maintenance and control optimization. Current implementations span radio access network (RAN) optimization \cite{Manalastas2024Simulators}, network planning \cite{Masaracchia2023OpenRAN}, edge computing orchestration \cite{Plageras2022DigitalTwinsIIoT}, and spectrum management \cite{NguyenKha2026DigitalTwinDSS}. Recent advances leverage generative AI including large language models and diffusion models for network optimization~\cite{li2025generative}, differentiable ray tracing for learning radio propagation environments~\cite{hoydis2024learning}, and foundation models pretrained on massive wireless channel datasets~\cite{alikhani2024lwm} demonstrating significant improvements in channel prediction accuracy.
However, these DT frameworks \cite{Manalastas2024Simulators,Masaracchia2023OpenRAN,Plageras2022DigitalTwinsIIoT,NguyenKha2026DigitalTwinDSS,khan2022digital, li2025generative, haider2025llmdt,hoydis2024learning,alikhani2024lwm}  built on statistical AI solutions do not reason about the world. They receive sensor measurements, update a statistical world model, and compute control and communication policies. But they do so reactively, without any mechanism to assess the reliability of their own predictions, identify what they do not know, or direct sensing resources to fill knowledge gaps. This passive relationship with the physical world creates a fundamental \emph{epistemic limitation}, which means the twin cannot identify what it does not know. When the physical AI network enters an unfamiliar regime, perhaps due to an unseen environmental change, or adversarial source, the twin will continue to make predictions based on its existing model, unaware that these predictions are unreliable,  thereby hindering the safe operation of physical AI systems.
Second, current DT architectures treat sensing as an external data source or some side information rather than manipulating the sensing resources to reduce epistemic uncertainty. Instead, physical AI systems require DTs that actively control sensing resources to reduce epistemic uncertainty, directing the network to acquire causally relevant information needed for reliable decision-making under uncertainty.
%In a 6G network with integrated sensing and communication (ISAC), the ability to actively probe the environment by directing beams to gather causally relevant information for the twin to perform inference represents a powerful capability.  

\subsubsection{Limitation 2: Absence of Compositional Guarantees for Heterogeneous Agents}
Physical AI systems comprise heterogeneous agents with fundamentally incompatible internal semantic representations about the world. In such a system, physical AI coordination requires agents to fuse complementary information from multiple neighbors to build a more complete picture of the shared environment than any single agent can achieve alone.  For example, consider a disaster response scenario via collaborative inference using ground robots and vehicles, aerial drones, and human extended reality (XR) user. Here, ground robots maintain occupancy grid maps from LiDAR, aerial drones build 3D environment models from cameras, vehicles track dynamic objects via radar, and human operators reason through natural language and visual interfaces. When the ground robot attempts to fuse the information received from other agents to update its world model, conflicts arise between representations that describe the same physical reality in incompatible ways. These representational conflicts propagate directly into the robot's action decisions, resulting in plans that appear individually rational but prove collectively catastrophic when executed in the shared physical environment. Current DT approaches, that use data-driven multimodal data fusion methods \cite{Salehi2022DeepLearningVehicular}, provide no formal mechanism to resolve such conflicts while preserving the causal and geometric structure that each agent relies upon for correct decision-making, making scalable heterogeneous physical AI coordination impossible without compositional guarantees.

\subsubsection{Limitation 3: Data-Driven Models Without Causal Understanding}

The dominant methodology for DT modeling employs data-driven approaches including generative AI  trained on historical data to predict network and physical agent behavior \cite{alikhani2024lwm,Dong2019DeepLearningDigitalTwin,Groshev2021DigitalTwinAI}. While powerful for capturing statistical correlations, these approaches exhibit several limitations.
First, they require massive training datasets covering all scenarios the physical AI network might encounter. But physical AI environments are inherently non-stationary and may encounter unexpected scenarios that may occur due to emergent interactions among the agents. No historical dataset can anticipate all futures. Moreover, neural models extrapolate poorly beyond their training distribution since they learn purely statistical associations between inputs and outputs \cite{thomas2023causal}. For example, in the disaster response scenario, a correlational model accurately learns that robot positions could be correlated with channel quality degradation. But  it cannot answer: ``If I command this robot to move behind the metal shelving, will the link fail before it completes its task?" Such counterfactual reasoning about physical interventions requires causal understanding that correlation-based models lack.
Finally, these models provide point predictions without principled uncertainty quantification. A DT might predict that a robot will reach its target location in $4$ seconds, but how reliable is this prediction? How does uncertainty about obstacle positions, floor stability, or neighboring agent trajectories propagate into uncertainty about mission success? Without answers to these questions that require understanding how the data is generated, the DT cannot guide risk-aware decision making and hence are \emph{not trustworthy in physical AI applications}. These limitations extend to recent agentic AI
frameworks that coordinate multiple LLM-based agents
through message passing over reliable wireless
channels \cite{Tran2025MultiAgentLLM}. While agentic AI has gained significant
attention for task decomposition and collaborative
problem solving, these frameworks rely on statistical models
that lack causal understanding of the physical world
and coordinate heuristically without evaluating
whether a given exchange genuinely improves the
receiver's uncertainty about the physical environment.
Physical AI coordination requires agents that reason
about hidden physical states through causal world
models, justify every inter-agent transmission by its
cognitive value to the receiver, and formally quantify
when coordination genuinely exceeds independent
operation rather than merely consuming resources.

\subsubsection{Limitation 4: Data Synchronization via Centralized Processing}
Current physical-virtual twin synchronization schemes assume centralized processing in which raw sensing data is communicated to a central location, processed there, and then decisions are distributed back. This architecture was appropriate when computation was expensive and endpoints were passive sensors. Physical AI coordination fundamentally inverts this model. Autonomous robots, drones, and vehicles must perform distributed inference about their environment and coordinate through belief exchange, not centralized data aggregation. Consider a team of ground robots navigating a collapsed building. The traditional centralized approach requires each robot to stream its full LiDAR point cloud to a central server for processing, which then communicates navigation decisions back. Under damaged infrastructure where communication links are intermittent and latency is unpredictable, this leads to delayed decisions that can hinder the safe operation of physical AI systems. With physical AI agents, each robot HDT must maintain a world model that predicts obstacle positions, assesses structural stability, and infers neighboring agents’ trajectories through theory-of-mind reasoning. These HDTs can further coordinate by exchanging inferred beliefs such as ``northeast corridor has $90$ percent probability of collapse within $60$ seconds" or ``agent $3$ trajectory will intersect this position in 200ms" . Such information  enables distributed collision avoidance and mission planning while consuming a fraction of the bandwidth that raw sensor streaming would require. Centralized processing cannot support this because the round-trip latency to a central server exceeds the time available for safety-critical decisions, and because link failures under damaged infrastructure make centralized coordination inherently fragile. What is required instead is a network that provides a shared spatiotemporal context among distributed agents by ensuring that belief updates, sensing decisions, and control actions are synchronized so that physically coupled agents can act as a single coordinated system.

%Consider base stations coordinating handover for autonomous vehicles. The traditional approach transmits raw CSI, potentially hundreds of complex coefficients per vehicle, to a central controller for processing. With physical AI agents, each base station HDT and vehicle HDT must maintain its own world model of mobility dynamics, channel evolution, and collision constraints. These HDTs coordinate via exchanging inferred beliefs such as ``vehicle trajectory causally predicts handover necessity in 100 ms with 85 percent confidence" or ``interference will spike when detected obstacle causes beamforming adaptation." These semantic information convey equivalent coordination value while consuming orders of magnitude less bandwidth than raw measurement streams.
%The Shannon paradigm optimized for faithful reproduction because endpoints could not reason, only measure and forward to centralized processors. Physical AI agents are instead autonomous reasoners maintaining causal world models and executing safety-critical decisions locally or at the edge. %Coordination requires transmitting beliefs that enable distributed causal reasoning, not raw data for centralized reconstruction. 
%Here, the fundamental bottleneck to scaling networks to thousands of heterogeneous physical AI agents under real-time constraints is the need to synchronize using limited semantic information.

\vspace{-2mm}\subsection{The Case for Holonic Digital Twins}
\label{three_shifts}
To address the above limitations of traditional DTs, we envision HDTs as cognitive agents that live within and continuously refine causal world models of their physical environment. Unlike the statistical world models of passive DTs that are updated periodically from sensor streams, a HDT's world model is the active substrate through which agents perceive hidden states, simulates the consequences of actions before executing them, resolves uncertainty through targeted sensing, and coordinates with neighboring agents through belief exchange. Hence, a cognitive DT uses its world model as a reasoning engine for intervention, actively participating in the perception-decision-action loop of its physical counterpart rather than observing it from the outside. Realizing this vision requires three foundational shifts in how DTs are designed, deployed, and coordinated across the network. %To address the above limitations of traditional DTs, we envision HDTs as cognitive agents that actively reason about their environment, direct sensing resources to reduce uncertainty, coordinate with peers (which include other AI agents as well as network infrastructure components) through belief exchanges, and pursue goals on behalf of their operators.
%This reconceptualization requires three foundational shifts:

1) \emph{From observation to causal inference:} The HDT should  actively maintain generative causal models $P(s_t|\mu_t, a_t)$ that explain how hidden environmental states $\mu_t$ at time $t$, which may include channel conditions, physical obstacles, neighboring agent intentions, and interference sources, causally give rise to observed measurements $s_t$ which can be cross-layer data such as channel estimates, LiDAR point clouds, sensor readings, upon performing an action $a_t$. This requires probabilistic inference over belief distributions $Q(\mu_t|s_{1:t})$ that quantify both \emph{aleatoric uncertainty} inherent in stochastic physical processes and \emph{epistemic uncertainty}  arising from incomplete observations or model misspecification \cite{hullermeier2021aleatoric}. Critically, these beliefs must support interventional queries via do-calculus: given current beliefs, the HDT should predict $P(\mu_{t+r}|do(a), s_{1:t})$ which is the causal effect of potential actions on future states. Furthermore, this should enable counterfactual reasoning essential for tackling unforeseen scenarios, performing theory of mind reasoning to infer other HDT beliefs and actions and ensuring accurate control decisions even under adversity.

2) \emph{From centralized to collective intelligence:} Physical AI coordination requires distributed frameworks where agents compose beliefs from heterogeneous neighbors while preserving causal consistency, enabling collective inference that is both bandwidth-efficient and compatible with safety-critical latency constraints. While distributed robotics frameworks for multi-robot
SLAM, consensus-based control, and decentralized task
allocation~\cite{Rizk2019MultiRobotSurvey} enable agents to
coordinate without a central server, they exchange
raw measurements and point state estimates without
evaluating how the transmitted information reduces
uncertainty at the receiver or whether it is causally
relevant to the receiver's next decision. Contrary to this, HDTs must determine which agents
are causally relevant to each other's decision-making
and prioritize transmissions that maximize cognitive
value at the receiver. Based on this causal structure, agents must dynamically establish coordination relationships, wherein they exchange semantic beliefs when causal coupling justifies communication overhead or operate independently when causal influences are negligible. This measured, adaptive HDT integration \emph{optimizes the tradeoff between collective intelligence and real-time bandwidth and latency constraints}.

3) \emph{From data to meaning to cognitive value:} Communication between HDTs should transmit beliefs rather than raw data, since what a neighboring agent requires is not a replica of another agent’s sensor measurements, but an update to its own world model about hidden states it cannot directly observe. For example, a ground robot does not require a drone’s raw camera frames. Instead, it requires the drone’s inferred belief about structural stability in the region it is about to enter, along with the associated uncertainty so that it can appropriately weight this information against its own observations. The network
must therefore evaluate each transmission by its
cognitive value, how much the transmitted belief
reduces the receiver's uncertainty about causally
relevant hidden states and improves its next physical
action, rather than by the number of bits it carries. This belief-centric communication enables each receiving agent to update its causal world model with complementary information from neighbors, thereby resolving collective uncertainty that no single agent can resolve independently.
%Communication between DTs should transmit beliefs, uncertainties, and causal predictions rather than raw data. While semantic communication compresses information to preserve task-relevant content, HDTs exchange probabilistic beliefs that enable distributed causal inference and thereby emergence of collective intelligence. In particular, semantic communication asks ``what task-relevant features should I transmit?" and optimizes encoding accordingly. Contrary to this, HDTs ask ``what beliefs enable coordinated causal reasoning?" and exchange posterior distributions, uncertainty quantifications, and causal model updates. 
%This belief-centric communication enables receiving agents to update their own causal world models and maintain semantic consistency across heterogeneous representations, going beyond task-specific compression to support compositional collective intelligence.

We provide a detailed comparison of the capabilities enabled by HDT-Net vs traditional DTs in Table~\ref{tab:comparison}. The foundational four pillars of HDT-Net form a single computational chain across sensing, communication, computing and control domains
rather than independent contributions. MBs determine \emph{with whom} each HDT must
coordinate based on causal relevance to its mission.
Active inference determines \emph{what} to transmit
by evaluating the cognitive value of each potential
belief exchange. Category theory determines
\emph{how} to translate beliefs and understand each other at the interface
between heterogeneous HDTs whose internal
world model representations may be incompatible. Spatiotemporal
integrated information $\Phi$ determines
\emph{whether} the resulting coordination leads to
 collective intelligence that justifies its
communication cost, feeding back into communication topology
adaptation when it does not. The following sections
present the HDT-Net architecture for 6G networks,
developed upon these pillars. %The following sections present the HDT-Net architecture for 6G networks, developed upon causal Markov blankets, active inference, and IIT as the key theoretical pillars enabling these shifts.

\begin{table*}[t]
\centering
\caption{Comparison of Traditional Digital Twins and Holonic Digital Twins (HDT-Nets).
%\cmark~=~Supported, \xmark~=~Not Supported, \pmark~=~Limited/Partial.
}
\label{tab:comparison}
\renewcommand{\arraystretch}{1.25}
\begin{tabular}{lcc}
\toprule
\rowcolor{headerblue}
\textcolor{white}{\textbf{Capability}} &
\textcolor{white}{\textbf{Traditional DT}} &
\textcolor{white}{\textbf{HDT-Net (Proposed)}} \\
\midrule
 
\rowcolor{groupblue}
\multicolumn{3}{l}{\textbf{\textcolor{headerblue}{World Model \& Reasoning}}} \\
World model for prediction                                      & \cmark & \cmark \\
\rowcolor{rowalt}
Causal world model (physics, geometry, causality)              & \xmark & \cmark \\
Epistemic uncertainty quantification                            & \xmark & \cmark \\
\rowcolor{rowalt}
Counterfactual reasoning via do-calculus                        & \xmark & \cmark \\
 
\midrule
\rowcolor{groupblue}
\multicolumn{3}{l}{\textbf{\textcolor{headerblue}{Sensing \& Perception}}} \\
Passive sensing from data streams                               & \cmark & \cmark \\
\rowcolor{rowalt}
Active sensing to reduce epistemic uncertainty                  & \xmark & \cmark \\
Long-horizon planning under uncertainty                         & \xmark & \cmark \\
 
\midrule
\rowcolor{groupblue}
\multicolumn{3}{l}{\textbf{\textcolor{headerblue}{Coordination \& Composition}}} \\
\rowcolor{rowalt}
Centralized data aggregation                                    & \cmark & \xmark \\
\rowcolor{rowalt}
Compositional fusion with semantic consistency guarantees       & \xmark & \cmark \\
Theory of mind for neighboring agents                           & \xmark & \cmark \\
 
\midrule
\rowcolor{groupblue}
\multicolumn{3}{l}{\textbf{\textcolor{headerblue}{Network-Level Intelligence}}} \\
\rowcolor{rowalt}
Physical-layer KPI optimization                                 & \cmark & \cmark \\
Cognitive value-driven transmission decisions                   & \xmark & \cmark \\
\rowcolor{rowalt}
Collective intelligence measurement ($\Phi$)                    & \xmark & \cmark \\
Adaptive coordination topology                                  & \xmark & \cmark \\
\rowcolor{rowalt}
Shared spatiotemporal context for physical AI swarms                        & \xmark & \cmark \\
 
\bottomrule
\end{tabular}
\vspace{-3mm}\end{table*}

\vspace{-2mm}\section{Causal Markov Blankets for Compositional
Physical AI Coordination}
\label{sec:markov_blankets}

Enabling HDTs for physical AI coordination requires addressing a fundamental design challenge absent from traditional wireless networks. This involves determining which devices should reason jointly versus independently, at what level of abstraction, and with what functional boundaries, based on the causal structure of the mission rather than on fixed network architecture. These decisions cannot be predetermined by network
architecture, protocol layers, or physical proximity. They depend on the causal structure of the mission,
specifically how agents influence each other through
shared spectrum, environment, or goals, how sensing
observations provide information about hidden states,
and how control decisions causally affect mission
outcomes.
%In conventional networks, functional decomposition is predetermined, where physical devices generate and consume data, base stations manage radio resources, core networks handle routing and mobility, and edge servers provide computation. But for physical AI coordination, such fixed architectures becomes a coordination bottleneck to achieve real-time inference. This naturally leads to an optimal decomposition that depends on causal structure that describes how physical AI agents influence each other through shared spectrum or environment or goals, how sensing observations provide information about  hidden states, and how control decisions causally affect  mission outcomes.
Considering the disaster response example, the system must determine whether each ground robot requires its own reasoning twin, whether aerial drones should share a single twin or maintain independent ones, and whether the network edge should host regional twins coordinating multiple robots. These decisions depend not on hardware boundaries but on causal dependencies. Robots operating in separated building sections have minimal causal coupling and should maintain independent twins, while robots navigating shared debris fields have strong causal interdependencies requiring coordinated twins that reason jointly about collision avoidance and structural stability.
\emph{Causal MBs} formalize this partitioning by defining agent communication boundaries based on how information causally flows through the cyber-physical system. In other words, MBs classify physical AI agents in terms of how they can interact efficiently in terms of communication, sensing and computing resources, with their environment via wireless network \cite{BeckRamstead2025DynamicMB}. 
%For physical AI networks spanning warehouses with hundreds of mobile robots, disaster zones with heterogeneous ground and aerial platforms, or smart cities with millions of vehicles and IoT devices, centralized coordination leads to huge communication overhead and impact the real-time operation of mission critical services. Autonomous agents enable scalable self-organization, fault tolerance when individual components fail, and adaptability to rapidly changing physical environments.
%A MB defines the minimal set of variables that shield an agent's internal states from the rest of the world, which is the boundary through which all information flows. %By extending this concept from statistical dependence to causal structure, we obtain a formal answer to network partitioning that enables the three foundational shifts Section~\ref{three_shifts} identified.
We begin by formalizing the MB concept, then extend it to HDTs, and finally show how this partitions wireless networks into compositional cognitive agents.

\vspace{-4mm}\subsection{Dynamic MB Partition Across Sensing, Communication, and Control Domains}

We consider the generative world model of a physical AI agent as a joint distribution over all sensing, communication, and control variables in its environment, represented as a directed acyclic graph (DAG) that encodes the causal dependencies among these variables, hence forming a causal Bayesian network. \emph{Sensory states} $s
$ comprise multimodal observations flowing into the agent across all domains. %Sensing observations may include LiDAR point clouds, radar returns, and channel sounding measurements. Communication observations include received signal strength indicators, channel state information from pilot signals, and semantic messages from neighboring agents. Control observations include state feedback from actuated systems such as robot joint angles and vehicle velocities, task completion acknowledgments, and safety alerts.
\emph{Active states} $a$ are communication and control actions flowing out from the agent to affect the physical and network environment across domains. %Sensing actions include radar beam directions, measurement dwell times, and spectrum sensing schedules. Communication actions include transmit power levels, beamforming vectors, resource block allocations, and handover commands. Control actions include trajectory commands to autonomous vehicles, motion planning waypoints, task assignments to robots, and safety intervention commands. 
\emph{Internal states} $\mu$ are hidden variables shielded from direct external influence, representing the agent's beliefs maintained through cross-domain fusion. Examples of distinct states across sensing, communication and control are detailed in Table~\ref{tab:markov_blanket_partition}. The internal states $\mu$ are updated through sensory observations but never directly observed themselves, representing what the agent believes is true about the integrated sensing-communication-control environment rather than ground truth.  

An MB of a variable $X$ in a causal Bayesian network is the minimal set of variables that renders $X$ conditionally independent of all other variables. Here, 
$X$ represents a state variable of interest such as an agent's physical configuration, a DT's internal belief, an environment condition, or a network resource parameters. The MB comprises parents $\text{Pa}(X)
$ representing direct causal drivers of 
$X$ such as control inputs, sensory observations, or neighboring agents' actions; children $\text{Ch}(X)$
 representing variables directly caused by 
$X$ such as measured observations, communication quality metrics, or downstream control decisions; and co-parents $\text{CoP}(X)
$ representing other causal factors that influence 
$X$'s effects, such as environmental conditions affecting multiple agents simultaneously or other agents' states that create coupling through shared resources. The MB is therefore
$
\text{MB}(X) = \text{Pa}(X) \cup \text{Ch}(X) \cup \text{CoP}(X)$.
%The MB defines a \emph{boundary} around $X$ such that everything $X$ needs to know about the world flows through this boundary. 
More formally:
\begin{equation}
P(X | \text{MB}(X), W \setminus \text{MB}(X)) = P(X | \text{MB}(X)),
\label{eq:conditional_independence}
\end{equation}
where $W$ represents all variables in the network. %Given its MB, $X$ is conditionally independent of everything else.
Since the MB is defined on a causal DAG, it enables \emph{interventional and counterfactual reasoning} over sensing, communication, and control variables. This allows each HDT to answer queries such as ``If I direct this sensing beam toward the northeast corridor, allocate additional bandwidth to this agent, or command this robot to halt, how would my belief about the physical world change?". Hence, this enables predicting the causal consequence of actions on the agent's world model before they are executed in the physical environment. For an intervention $\text{do}(A)$ on variables outside $\text{MB}(X)
$, with $X\in \{s,a\}$: $
P(X | \text{do}(A), \text{MB}(X)) = P(X | \text{MB}(X))$.  This means that variables outside the MB have no causal influence on the agent's internal beliefs. Consequently, the agent need not sense, communicate with, or respond to any entity outside its MB, enabling principled pruning of sensing resources, communication links, and coordination relationships to only those that causally matter for the agent's mission. This interventional property distinguishes causal reasoning from mere correlation and proves critical for efficient real-time autonomous decision-making under resource constraints. %In physical AI systems, the MB naturally partitions into components that span the integrated sensing, communication and control domains where physical AI agents operate, creating cross-domain tradeoffs. Moreover, the optimal partitions are time varying as the environment and task dynamics changes. This necessitates that the agent boundaries must be dynamically adjusted to utilize communication, sensing, and computing resources efficiently. 
However, traditional MB formulations assume static partitions where distinct network parameters maintain fixed roles as internal, boundary, or external states. This restriction fails for physical AI coordination under mobile agents, highly dynamic wireless channels, and time varying coordination requirements or tasks. Recent work on dynamic MB detection~\cite{BeckRamstead2025DynamicMB} addresses this limitation by allowing blanket elements to transition between roles over time while preserving conditional independence structure.
For HDT-Net, dynamic blankets enable causality-aware integration of sensing, communication, and control. In practice, each HDT tracks its blanket boundaries
by periodically evaluating the conditional mutual
information between its internal states and candidate
boundary variables using local observations. When
causal coupling with a neighboring agent increases,
for instance two robots converging on a shared
corridor, the neighbor's states enter the blanket
and a coordination link is established. When coupling
decreases, those states exit the blanket and the link
is released. The cognitive value metric defined in Section~\ref{sec:active_inference}
provides a computationally efficient proxy for this
adaptation. When the cognitive value of transmissions
across an existing link drops below the communication
cost, the causal coupling justifying that link has
weakened, triggering blanket contraction. Conversely,
when an HDT's expected free energy remains high
despite local observations, unresolved uncertainty
about states outside its current blanket triggers
exploration of new coordination links. The active
inference loop thereby drives MB adaptation as a
byproduct of free energy minimization rather than
requiring a separate blanket detection algorithm.%This is achieved via dynamically mapping sensing states as consisting of sensor observations that causally influence coordination decisions. Beliefs that must be exchanged across wireless links depend on the causal relevance of the information for the receiving agent's mission, the channel conditions among agents, and whether communication latency threatens control stability. %First, as physical AI agents navigate through environments, the set of wireless links constituting an agent's sensory states adapts based on channel quality and proximity. For example, an agent entering a building while transitioning from outdoor base station connectivity to indoor small cell association must dynamically reconfigure its MB boundary. Second, coordination scope adapts to mission requirements. During independent operation, agent blankets remain separate with minimal cross-boundary information flow. When collaborative tasks emerge requiring tight coordination, blankets merge through shared boundary states enabling belief exchange. Third, semantic content transmission adapts to causal relevance. Boundary states transmit only information affecting coordinated decision-making, filtering observations lacking cross-agent causal influence.
\begin{table*}[t]
\centering
\caption{MB Partition for Cyber-Physical DTs in Integrated Sensing, Communication, and Control Systems}
\label{tab:markov_blanket_partition}
\vspace{-2mm}\begin{tabular}{lp{0.12\textwidth}p{0.65\textwidth}}
\hline
\textbf{Component} & \textbf{Domain} & \textbf{Examples} \\
\hline
\multirow{4}{*}{\textbf{Sensory States}} & Communication & Channel state information (CSI) estimates from pilots, received signal strength indicator (RSSI) measurements from users, ACK/NACK feedback from transmissions \\
& Sensing & Radar returns for object detection, LiDAR point clouds for environment mapping, Internet of Things (IoT) sensor readings (temperature, position, velocity) \\
& Control & State feedback from controlled agents (vehicle positions, drone trajectories, robot joint angles), task completion acknowledgments, safety alerts \\
& Cross-domain & Traffic arrival statistics, interference measurements, quality-of-service metrics \\
\hline
\multirow{4}{*}{\textbf{Active States}} & Communication & Transmit power levels, beam configurations, resource block allocations, handover commands \\
& Sensing & Sensing beam directions, radar waveform parameters, measurement scheduling decisions \\
& Control & Trajectory commands to autonomous vehicles, task assignments to robots, motion planning waypoints, safety intervention commands \\
& Cross-domain & Joint sensing-communication beam patterns, resource allocation balancing communication and control latency \\
\hline
\multirow{4}{*}{\textbf{Internal States}} & Communication & Estimated channel matrices, user mobility models, interference pattern predictions, traffic demand forecasts \\
& Sensing & Environment maps (obstacles, dynamic objects), target tracking filters, clutter statistics \\
& Control & Agent state estimates (positions, velocities, intentions), task progress models, safety constraint representations \\
& Integrated & Joint channel-environment models (how obstacles affect propagation), communication-aware motion planning (network topology constraints on control), uncertainty representations across all domains \\
\hline
\end{tabular}
\vspace{-3mm}\end{table*}
%The causal structure encodes how actions influence observations across domains: increasing sensing beam dwell time improves object localization (sensing) but reduces communication resources (cross-domain tradeoff); transmitting control commands consumes communication capacity while enabling actuation (communication-control coupling); autonomous vehicle trajectories affect channel conditions through mobility (control-communication feedback).

\vspace{-4mm}\subsection{Compositional MBs via Categorical Structures}
Individual MBs define single-agent boundaries, but scalable physical AI systems require composing hundreds or thousands of agents performing distributed inference across multiple spatial and temporal scales. For example, in the disaster response scenario, aerial drones with wide-area thermal imaging must fuse observations with ground robots using LiDAR for performing high-resolution structural assessment of collapsed buildings to maximize mission effectiveness. These multi-scale coordination scenarios require compositional MBs where smaller HDTs combine through belief exchange to form larger cognitive units exhibiting collective intelligence.  This capability enables hierarchical, cross-scale reasoning and coordination, allowing the system to integrate complementary sensing modalities, resolve partial and uncertain observations, and make globally consistent decisions that no individual agent could achieve alone. Compositional structure ensures that blankets at different hierarchical levels preserve conditional independence while enabling coordinated inference. %After identifying the constituent blankets, agents must decide how to coordinate under wireless bandwidth constraints and latency requirements for real-time control. 
However, efficient HDT composition is hindered by \emph{semantic misalignment}, where heterogeneous agents with different sensor modalities and world models produce beliefs that conflict rather than fuse coherently when combined. %Second, \emph{cross-layer inconsistency}: decisions optimized independently at the networking layers via distinct AI models contradict each other when composed. %Third, \emph{vendor incompatibility}: different HDT implementations from multiple vendors cannot exchange beliefs or coordinate actions without manually constructed conversion layers for each vendor combination \cite{SanaStrinati2023SemanticEQ}, which introduce errors as semantic meaning degrades through successive transformations between incompatible representations.
  Traditional AI systems operate as closed black boxes where internal representations and reasoning processes remain opaque and incompatible across vendors and AI agents. Successful coordination  requires heterogeneous agents to align their world models and exchange beliefs despite incompatible internal representations and implementations. %Compositional MBs via categorical structures provide the mathematical foundation for open intelligence architectures. %Each agent's MB boundary specifies the probability distributions it transmits and their causal dependencies, enabling receiving agents to correctly incorporate external beliefs into their own inference regardless of internal algorithmic differences. 
 MB boundaries define precisely what information crosses agent boundaries  versus what remains proprietary (agent-specific internal representations and algorithms).  Category theory further formalizes how abstraction levels compose with guaranteed consistency, and how different agent world model implementations can interoperate through standardized interfaces.

\subsubsection{Category Theory Primer for HDTs}

We formalize HDT composition through the category $\mathbf{NetHDT}$. Objects in this category are HDT states $(s, a, \mu)$, whereas morphisms are structure-preserving transformations between states which includes belief updates from new observations, state transitions from executed actions, and coordination protocols exchanging semantic information across MB boundaries. Morphism composition respects \emph{temporal causality}: where belief update $f: A \to B$ followed by state transition $g: B \to C$ yields the composed operation $g \circ f: A \to C$, which preserves the temporal ordering among the events. Each state possesses an identity morphism $\text{id}_A: A \to A
$ representing no change. This structure satisfies associativity $(h \circ g) \circ f = h \circ (g \circ f)$
 and identity laws $\text{id}_B \circ f = f = f \circ \text{id}_A
$, guaranteeing that the order of grouping operations does not affect outcomes while preserving causal temporal ordering, critical for distributed multi-agent coordination where updates occur asynchronously.

\subsubsection{Functorial Mappings Between Abstraction Levels}
Physical AI coordination requires agents to reason
at multiple levels simultaneously. A ground robot
must reason about centimeter-level obstacle geometry,
while the edge server coordinating fifty such robots
must reason about zone-level congestion and task
allocation. A HDT must also predict what its neighbors
believe about the shared environment, even when those
neighbors use entirely different sensor modalities and
world model representations. These multi-level
reasoning capabilities require mappings between
abstraction levels that preserve the causal
relationships agents rely upon for correct
decision-making.
Category theory formalizes such
mappings as functors: a functor
$F: \mathcal{C} \to \mathcal{D}$ maps objects and
morphisms from category $\mathcal{C}$ to category
$\mathcal{D}$ while preserving composition
$F(g \circ f) = F(g) \circ F(f)$ and identity
$F(\text{id}_A) = \text{id}_{F(A)}$.  The preservation
means that a functor cannot
introduce causal relationships that do not exist in
the source, nor destroy causal relationships that do.
For physical AI networks, we define three functors
that enable multi-level reasoning.

The \emph{physical-to-virtual functor}
$\Phi_{\text{DT}}$ maps physical world entities and
events to their DT world model representations. This is the mapping that turns the
real world into the twin's internal model, formalized
as the causal generative model in Section~IV.  Crucially, this functor preserves causal structure, ensuring that if physical event $
A$ causes physical event 
$B$ or object $A$ is related to object $B$, then the digital representation $\Phi_{\text{DT}}(A)$
 maintains the same causal relationship to $\Phi_{\text{DT}}(B)$ in the twin's world model, enabling valid counterfactual reasoning about interventions.
The \emph{abstraction functor} $\Phi_{\text{abs}}$ maps detailed HDT states to coarse-grained representations (via a many to one mapping) for hierarchical reasoning. For example, instantaneous LiDAR point clouds from individual ground robots become aggregate obstacle maps at the edge server HDT, and per-robot collision avoidance decisions become coordinated formation plans at the team level. This enables an HDT at the edge server to coordinate  in real-time hundreds of agents through low-dimensional causal abstractions capturing coordination-relevant information, rather than using full dimensional observations which wastes bandwidth and time. %These abstractions provide sufficient statistics for determining which zones require bandwidth, where coordination bottlenecks emerge, and how task priorities shift, without processing high-dimensional full state detail from each individual robot.
The \emph{theory of mind functor}
$\Phi_{\text{ToM}}$ formalizes how a HDT models what
a neighboring agent believes about the shared
environment. In particular, $\Phi_{\text{ToM}}
$ maps the internal belief state $\mu_i$ of agent $
i$ to a prediction $\hat{\mu}_j$ of what agent 
$j$ believes about the same physical environment, despite the two agents using incompatible world model representations. Without this capability, an agent sending a belief
update to a neighbor has no way of knowing whether
that update is redundant, surprising, or contradictory
from the neighbor's perspective. Crucially, theory of
mind is what makes cognitive value of information computable. To
evaluate how much a message will reduce the receiver's
expected free energy, the sender must first estimate
what the receiver currently believes and how the
message would alter that belief. Without
$\Phi_{\text{ToM}}$, the cognitive value
in~\eqref{eq:cog_value} cannot be evaluated, and
transmission decisions fall back to heuristic
scheduling that treats all messages as equally
valuable. By assessing the cognitive value of a potential
transmission via theory of mind before committing
communication resources, each HDT directs bandwidth
toward messages that most improve collective
physical AI performance. In~\cite{Wang2026MetaMind}, we showed
that agents can learn such metacognitive capabilities
in a self-supervised manner, generalizing from
first-person self-reflective inference to third-person
reasoning about other agents' goals and beliefs
through analogical reasoning. $\Phi_{\text{ToM}}$
provides the categorical formalization of this
capability, mapping the internal belief
state~$\mu_i$ of agent~$i$ to a prediction
$\hat{\mu}_j$ of what agent~$j$ believes, despite
the two agents using incompatible world model
representations.

\subsubsection{Compositional HDTs via Pullbacks}

\begin{figure}[t]
\centering
\begin{tikzpicture}[scale=1.2, >=stealth]
  % Nodes
  \node (DT12) at (0,2.5) {$\text{DT}_{1,2}$};
  \node (DT1) at (-2,0.5) {$\text{DT}_1$};
  \node (DT2) at (2,0.5) {$\text{DT}_2$};
  \node (Shared) at (0,-1) {Shared State};
  
  % Solid arrows
  \draw[->, thick] (DT12) -- (DT1) node[midway, above left] {$\pi_1$};
  \draw[->, thick] (DT12) -- (DT2) node[midway, above right] {$\pi_2$};
  \draw[->, thick] (DT1) -- (Shared) node[midway, left] {$f_1$};
  \draw[->, thick] (DT2) -- (Shared) node[midway, right] {$f_2$};
  
  % Dashed diagonal showing commutative property
  \draw[->, dashed, thick] (DT12) -- (Shared) node[midway, right, font=\small] {$f_1 \circ \pi_1 = f_2 \circ \pi_2$};
  
\end{tikzpicture}
\vspace{-3mm}\caption{\small Pullback diagram for composing digital twins. The composite $\text{DT}_{1,2}$ projects to individual DTs via $\pi_1$ and $\pi_2$, which then map to shared state via $f_1$ and $f_2$. The pullback property ensures $f_1 \circ \pi_1 = f_2 \circ \pi_2$, maintaining consistency on shared components.}
\label{fig:pullback}
\vspace{-6mm}\end{figure}
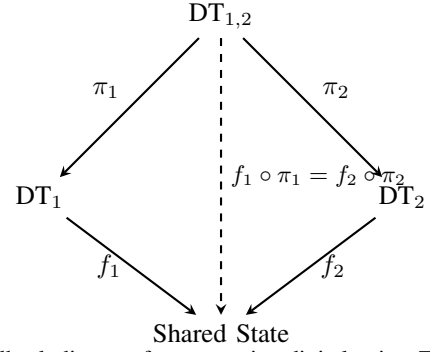

When multiple HDTs must coordinate through shared physical states or joint action constraints, we require a compositional operation that preserves each twin's individual MB structure while creating the joint structure needed for coordination. Category theory provides this through the pullback construction.
Consider two physical AI agents HDT$_1$ and HDT$_2$ with states $(s_1, a_1, \mu_1)$ and $(s_2, a_2, \mu_2)$ that must coordinate through a shared physical resource or constraint. For instance, an aerial drone and ground robot jointly estimating structural stability of a damaged building share observations of the same structure, or multiple autonomous vehicles navigating an intersection share the constraint that their trajectories must remain collision-free. The pullback HDT combines sensory states as $s_1 \times s_2$, constrains active states such that $(a_1, a_2)$ satisfy joint physical feasibility requirements, and represents joint beliefs $\mu_{1,2}$ that factor according to conditional independence:
$
P(\mu_1, \mu_2 | s_1, s_2) = P(\mu_1 | s_1, s_{\text{shared}}) \cdot P(\mu_2 | s_2, s_{\text{shared}})$,
where $s_{\text{shared}}$ represents observations of shared physical states that both twins observe or constraints both must satisfy. This is illustrated in Fig.~\ref{fig:pullback}. This factorization preserves each twin's individual MB by ensuring that beliefs about agent-specific local states remain conditionally independent given shared interface observations. This avoids unnecessary coupling that could lead to redundant information exchange and bandwidth wastage, while still ensuring coordination where physical causality demands it. While pullback formalizes composition of MBs, it does not alone guarantee \emph{semantic consistency}, which means that heterogeneous twins with different sensor modalities or world model representations must maintain compatible beliefs about shared physical states at their interface. This can be achieved via natural transformations.

\subsubsection{Natural Transformations for Multi-Agent Interoperability}
%The pullback composition developed in the previous subsection ensures conditional independence structure is preserved when HDTs coordinate. This assumes HDTs operate in compatible state spaces. However, physical AI deployments violate this assumption.  
Different HDT implementations encode the same physical phenomena in fundamentally different semantic representations, and hence different \emph{semantic languages}. For example, when a drone sends a belief update to a ground
robot, the drone's message is expressed in visual
embeddings while the robot reasons in geometric
structural models. Without a principled translation
mechanism, these incompatible representations produce
conflicting or meaningless updates when fused. Natural transformations provide the categorical machinery for translating between semantic languages while preserving meaning.
 Consider disaster response coordination where an aerial drone HDT represents building structural state using visual feature embeddings from learned neural encoders, while a ground robot HDT represents the same building using geometric structural models with explicit load-bearing wall positions and stress distributions. These are distinct semantic languages since vendor A's language has no explicit geometric primitives and encode damage as vectors in learned embedding space, while vendor B's language has no learned features, encoding damage as stress tensor fields over geometric primitives. Both observe the same damaged structure but use mutually incompatible semantic languages to represent internal beliefs. 

 Natural transformations between time-indexed functors
provide translations between these semantic languages.
We model the mission timeline as a category
$\mathbf{Time}$ whose objects are discrete time steps
$t = 0, 1, 2, \ldots$ and whose morphisms
$t \to t+1$ represent the causal passage of one
inference cycle. Each agent's semantic evolution is
then a functor from this timeline into its own
representation category:
$F_A: \mathbf{Time} \to \mathbf{Cat}_A$ maps each
time step to the drone's current visual embedding
state and each morphism to the belief update triggered
by new imagery, while
$F_B: \mathbf{Time} \to \mathbf{Cat}_B$ maps each
time step to the robot's current geometric model state
and each morphism to the structural model update from
new sensor data. 
 A natural transformation $\alpha: F_A \Rightarrow F_B
$ translates between semantic languages while preserving belief update semantics. When the drone updates its visual embedding beliefs with new imagery via functor $f_A$, this update must correspond to equivalent updates in the robot's geometric model via functor $f_B$. This semantic interoperability is achieved via natural transformation that ensures what is communicated over the air, i.e. $\alpha_{t+1}(f_A(s_t))$ ensures semantic alignment, thanks to its commutative property: $
\alpha_{t+1} \circ f_A = f_B \circ \alpha_t$.
This commutativity guarantees that translating then updating yields the same beliefs as updating then translating, enabling the drone to communicate ``northeast corner shows severe structural damage" in a form the robot can interpret for navigation planning. Heterogeneous physical AI agents thereby maintain coherent shared world models despite incompatible internal semantic languages, with natural transformations ensuring semantic consistency at MB boundaries.

In short, natural transformations enable \emph{heterogeneous agent interoperability} where agents using incompatible internal representations, exchange semantic information over wireless channels while  ensuring semantic interoperability despite different implementation choices. Pullback composition guarantees that independently developed HDTs \emph{compose through standardized wireless interfaces} with well-defined behavior rather than undefined interactions. Functorial abstraction enables \emph{hierarchical coordination} where edge servers reason over aggregated states while physical agents maintain fine-grained beliefs, with the agents partitioned across distinct MB levels in a way as to ensure real-time distributed inference.
Category theory hence provides the formal foundation ensuring HDTs can be built modularly, verified independently, and composed systematically through wireless networks while preserving semantic consistency.

\vspace{-3mm}\section{Compositional Active Inference and Emergent Collective Intelligence}

\label{sec:active_inference}

Causal MBs determine how the network partitions into autonomous reasoning agents and govern how causal information flows across agent boundaries. Herein, each blanket define the minimal set of sensing, communication and control variables a HDT must intervene upon and the beliefs that must be exchanged to perform accurate distributed inference without wasting communication resources.  However, blanket
structure alone does not prescribe how each HDT
should update its beliefs from incoming observations,
select actions across sensing, communication, and
control domains, or continuously refine its world model from
experience. Active inference provides this missing
optimization principle that enables collective intelligence among the networked HDTs within these boundaries \cite{Hashash2026ActiveInferenceScalingLaw}. Rather than implementing perception, planning, and control as separate engineered subsystems, each HDT executes a single computational objective called \emph{variational free energy minimization (VFE)}. The solution to VFE simultaneously produces belief updates about hidden world states, joint sensing, communication and control action selection that balances goal achievement against uncertainty reduction, and continuous learning that refines the causal generative model from experience.
 %Active inference provides the optimization framework for how HDTs perceive, act, and learn within the MB boundaries established in Section III. Physical AI networks, however, comprise many such agents coordinating through wireless links, creating challenges that single-agent active inference does not address. %Individual active inference processes must compose to enable inference when heterogeneous HDTs coordinate, requiring formal frameworks that preserve performance guarantees. 
%Physical AI agents exchanging semantics need mechanisms ensuring semantic consistency despite intermittent wireless links  and heterogeneous internal representations. Expected free energy minimization across multi-agent systems operating under communication constraints demands decomposition into individual objectives plus coordination terms that account for wireless resource consumption.
Physical AI systems, however, comprise many such agents
coordinating via  wireless links across distinct spatial and temporal scales. This creates challenges
that single-agent active inference does not address. To address these challenges, we first
show how active inference instantiates the perception-action
loop of physical AI agents within MB boundaries. We then
show that active inference has categorical structure,
enabling HDTs to compose via pullbacks with provable free
energy bounds. We next derive conditions under which networked HDTs, organized across hierarchical MB levels, collectively behave as a single active inference agent at each scale, showing that genuine collective intelligence emerges bottom-up from local free energy minimization through wireless belief exchange rather than from top-down centralized coordination. Finally, we introduce IIT as the measure that quantifies when this collective
intelligence exceeds independent operation, and derive
growth dynamics that govern how it evolves over mission
timescales.

\vspace{-3mm}\subsection{The Generative Model and Variational Inference}

At the heart of active inference lies a generative model encoding the causal structure of how hidden states give rise to observations. For physical AI coordination through wireless networks, this model must span integrated sensing, communication, and control domains.
The generative model factorizes as $
P(\tilde{s}, \tilde{\mu}, \pi) = P(\mu_0) \prod_{t=0}^{T} P(s_t|\mu_t) P(\mu_{t+1}|\mu_t, a_t) P(\pi)$,
where $\tilde{s}$ and $\tilde{\mu}$ denote sequences of observations and states, $a_t$ is the action at time 
$t$ determined by policy $\pi$, and the factors represent initial state distribution, observation likelihood capturing how physical and network conditions cause sensor measurements, state transition dynamics governing how actions causally affect future states, and policy prior encoding task preferences. %For physical AI networks, states $s_t$ include channel matrices, agent positions and velocities, obstacle configurations, and interference patterns. Observations $o_t$ comprise channel state information,  measurements from distinct sensor modalities, agent state feedback, and network quality metrics. Actions $a_t$ span communication (power, beamforming, resource allocation), sensing (beam directions, measurement scheduling), control (trajectories, task assignments), and cross-domain coordination balancing latency against stability.
\emph{Perception} corresponds to the HDT maintaining
approximate posterior beliefs $Q(\mu_t)$ that minimize
variational free energy:
\vspace{-2mm}\begin{equation}
    \mathcal{F} =
    \underbrace{{\mathrm{KL}}\bigl[Q(\mu_t)
        \,\|\, P(\mu_t | s_{<t}, a_{<t})\bigr]}_{\text{complexity}}
    - \underbrace{\mathbb{E}_Q[\ln P(s_t|\mu_t)]}_{\text{accuracy}}.
    \label{eq:vfe}
\vspace{-2mm}\end{equation}
The complexity term, measured as the Kullback-Leibler (KL) divergence between approximate and true posterior distribution about the beliefs $\mu_t$ is written as ${\mathrm{KL}}\bigl[Q(\mu_t)
        \,\|\, P(\mu_t | s_{<t}, a_{<t})\bigr] = \mathbb{E}_{Q(\mu_t)} \frac{\ln Q(\mu_t)}{\ln P(\mu_t | s_{<t}, a_{<t})} $, where $s_{<t}$ and
$a_{<t}$ denote all observations and actions prior
to time $t$ and $P(\mu_t | s_{<t}, a_{<t})$ is the true belief posterior. It penalizes unnecessary departure from what the HDT already knew about the physical world, while the accuracy term rewards beliefs that correctly predict incoming sensor observations. Minimizing~\eqref{eq:vfe} thus updates the HDT's world model to best explain incoming observations while staying anchored to its causal prior.
\emph{Action selection} corresponds to minimizing
expected free energy over future policies:
\vspace{-1mm}\begin{equation}
    \mathcal{G}(\pi) =
    \underbrace{\mathbb{E}_Q\bigl[D_{\mathrm{KL}}[Q(s_\tau|\pi)
        \,\|\, P(s_\tau)]\bigr]}_{\text{pragmatic value}}
    + \underbrace{\mathbb{E}_Q\bigl[H[P(s_\tau|\mu_\tau)]
        \bigr]}_{\text{epistemic value}},
    \label{eq:efe}
\end{equation}
where the pragmatic term drives the HDT toward preferred
future observations (goal achievement) and the epistemic
term drives actions that reduce ambiguity about hidden
states (uncertainty reduction). The critical property of
\eqref{eq:efe} is that exploration and exploitation are
not separately engineered. For example, a disaster response ground robot exploits
known collision-free paths when its belief about obstacle
positions is sharp, and explores cautiously when
uncertainty is high, with the balance determined entirely
by the ratio of pragmatic to epistemic cost at the current
belief state. A HDT that cannot yet explain its
observations well  prioritizes to spend more sensing  and communication  resources
 to gather information before committing to
goal-directed movement.

\emph{Learning} further closes the loop where as actions produce
new observations, the parameters of the generative
model are updated to improve future
predictions. This step refines both the observation likelihood
$P(s_t|\mu_t)$ and the state transition model
$P(\mu_{t+1}|\mu_t, a_t)$ from experience. This is not a
separate learning phase but a continuous process
co-occurring with perception and action. The same VFE
objective that drives belief updates also drives model
improvement, ensuring that learning is always performed with the goal of reducing surprise about the physical environment.

\subsubsection{Cognitive Value of a Transmission}
The perception-action-learning loop above governs a single
HDT. When multiple HDTs coordinate over a shared wireless network, real-time constraints, coupled with limited bandwidth, make it essential to quantify the extent to which a received message improves the receiver’s ability to sense, decide, and act. Herein, the expected free energy decomposition in~\eqref{eq:efe} provides
a principled formalization of the \emph{cognitive value} of information
. When agent~$i$ transmits a belief
message~$m_{i \to j}$ to agent~$j$, the cognitive value of that
transmission is the reduction in agent~$j$'s expected free energy
induced by incorporating the received belief into its posterior:
\vspace{-1mm}\begin{equation}
\mathcal{V}(m_{i \to j}) \triangleq
G_j\!\bigl(\pi^*_j \,\big|\, Q_j(\mu)\bigr)
- G_j\!\bigl(\pi^*_j \,\big|\, Q_j(\mu \,|\, m_{i \to j})\bigr),
\label{eq:cog_value}
\end{equation}
where $Q_j(\mathbf{s})$ is agent~$j$'s prior belief over hidden
physical states and $Q_j(\mathbf{s} \,|\, m_{i \to j})$ is the
updated posterior after receiving the message. Because the
EFE decomposes into epistemic and pragmatic terms,
cognitive value inherits the same decomposition:
\vspace{-3mm}\begin{equation}
\mathcal{V}(m_{i \to j}) =
\underbrace{\Delta \mathcal{V}_{\text{e}}(m_{i \to j})}_{\substack{\text{reduction in}\\
\text{state uncertainty}}}
+ \underbrace{\Delta \mathcal{V}_{\text{p}}(m_{i \to j})}_{\substack{\text{improvement in}\\
\text{goal alignment}}}.
\label{eq:cog_value_decomp}
\vspace{-1mm}\end{equation}
The epistemic component captures how much the received belief
reduces agent~$j$'s uncertainty about causally relevant hidden
states, while the pragmatic component captures how much
the updated belief enables agent~$j$ to select actions
closer to its preferred outcomes. For instance, a drone evaluating whether to transmit its structural assessment to a ground robot computes the cognitive value by estimating how much that assessment would reduce the robot's uncertainty about obstacle states governing its next navigation decision. Unlike classical value of
information~\cite{Howard1966InformationValue}, which measures
reduction in statistical uncertainty independent of the decision
context, cognitive value is inherently \emph{action-coupled}. A
message carries zero cognitive value if it reduces uncertainty
about states that are causally irrelevant to the receiver's
current policy selection, regardless of how much Shannon
information it contains.  A communication link is worth maintaining when
the aggregate cognitive value it carries exceeds its
communication cost, and should be pruned otherwise.
However, pairwise cognitive value exchange alone does
not guarantee that a network of HDTs constitutes a
coherent collective agent.%This connects directly to
\vspace{-4mm}\subsection{Networked HDTs and the Emergence of
Collective Intelligence via Hierarchical Twins}
 
%While Theorem~1 establishes the coordination benefit available to any pair of HDTs whose beliefs are statistically coupled through shared observations, statistical coupling alone does not guarantee that the collection of locally optimal solutions constitutes a coherent collective agent. 
Each HDT may correctly maximize the cognitive value
of its own transmissions and correctly minimize its
own VFE, yet the composed system lacks a principled free energy objective of its own. This makes it impossible to reason about collective behavior, plan hierarchically, or guarantee stability when the network topology changes. Resolving this requires showing that a network of HDTs, each minimizing its own VFE through wireless belief exchange, can itself be understood as a single active inference agent at the collective scale.
 
\subsubsection{From Individual HDTs to a Collective Agent}
 
Consider $N$ HDTs $\{\mathcal{R}_i\}_{i=1}^N$ deployed
across a physical AI network. When these HDTs coordinate
through wireless links, each twin's active states include
belief transmissions and each twin's sensory states include
received beliefs from neighbors. The inward-facing
components of the MB, $a_i^{\text{in}}$ and
$s_i^{\text{in}}$, mediate inter-HDT coupling, while the
outward-facing components, $a_i^{\text{out}}$ and
$s_i^{\text{out}}$, interface with the physical AI agent
and its environment:$
    a_i = (a_i^{\text{out}},\, a_i^{\text{in}}), \quad
    s_i = (s_i^{\text{out}},\, s_i^{\text{in}}).
$
With this decomposition, the causal chain through which
HDT~$i$ influences HDT~$j$ is fully determined:
\vspace{-2mm}\begin{equation}
    \mu_i \;\to\; a_i^{\text{in}}
           \;\to\; \eta_{ij}
           \;\to\; s_j^{\text{in}}
           \;\to\; \mu_j, \qquad i \neq j,
    \label{eq:coupling_chain}
\vspace{-2mm}\end{equation}
where $\eta_{ij}$ is the shared wireless channel state
mediating communication between twins $i$ and $j$.
Because all inter-HDT influence flows through the wireless interface rather than through direct coupling of internal states, each HDT's world model implementation remains proprietary behind its blanket boundary. Semantic translation between heterogeneous implementations is handled entirely at the boundary by the natural transformations of Section~III-C4, with no requirement for shared internal representations across agents. The explicit appearance of $\eta_{ij}$
 in~\eqref{eq:coupling_chain} means that wireless channel degradation enters the free energy accounting directly. A deteriorating link reduces the cognitive value of $s_j^{\text{in}}$, raises $\mathcal{F}_j$, and triggers either a link recovery action or a reduction of coordination scope, all as consequences of VFE minimization rather than as separately engineered layered protocol decisions.
 
The coupling structure~\eqref{eq:coupling_chain} produces
a \emph{blanket of blankets}: the outward-facing
components $\{s_i^{\text{out}}, a_i^{\text{out}}\}$
jointly constitute a collective MB that
screens the aggregate internal beliefs
$\mu_{\text{col}} = \{\mu_i, s_i^{\text{in}},
a_i^{\text{in}}\}_{i=1}^N$ from the collective external
environment $\eta_{col}$, defined as all physical world states outside the ensemble that influence it only through the outward-facing blanket components $\{s_i^{\text{out}}, a_i^{\text{out}}\}$. When two structural conditions
hold, a) timescale separation
($\tau_{\text{local}} \ll \tau_{\text{coord}}$,
reflecting that onboard inference at millisecond
timescales is faster than wireless coordination at
tens-of-millisecond timescales) and b) the existence of a
collective non-equilibrium steady-state density
$p(z_{\text{col}})$, the collective autonomous states
$\alpha_{\text{col}} = (\mu_{\text{col}},
a_{\text{col}})$ obey the gradient flow \cite{devries2026active}:
\vspace{-2mm}\begin{equation}
    \dot{\alpha}_{\text{col}} \;\propto\;
    \nabla_{\alpha_{\text{col}}} \log\,
    p(s_{\text{col}},\, a_{\text{col}},\,
      \mu_{\text{col}}).
    \label{eq:collective_flow}
\vspace{-2mm}\end{equation}
Equation~\eqref{eq:collective_flow} states that the HDT network,
viewed as a whole, obeys the same gradient flow that
governs any individual active inference agent. The
collective is not merely a set of agents that coordinate;
it is itself a cognitive agent that perceives, acts, and
learns at a higher abstraction level, without introducing
any principle beyond those governing individual HDTs.
This formally resolves both failure modes: the network
is not merely coordinating but genuinely reasoning
collectively, and the composed system has a
well-defined VFE objective of its own.
\subsubsection{Collective Intelligence as Recursive Free
Energy Minimization}
 
Because the collective agent minimizes its own VFE, it
inherits all properties of individual active inference
at the network level. The collective expected free energy
decomposes as:
\vspace{-2mm}\begin{equation}
    \mathcal{G}_{\text{col}}(\pi_{\text{col}}) =
    \sum_{i=1}^N \mathcal{G}_i(\pi_i)
    \;-\!\!\sum_{(i,j)\in\mathcal{E}}\!\!
        I(\mu_i;\mu_j \mid s_{\text{shared},ij})
    \;+\; \mathcal{C}_{\text{comm}},
    \label{eq:collective_efe}
\vspace{-2mm}\end{equation}
where $\mathcal{E}$ is the set of active wireless links
and $\mathcal{C}_{\text{comm}}$ captures the free energy
cost of wireless belief exchange which includes encoding distortion,
transmission energy, and scheduling delay.
Minimizing~\eqref{eq:collective_efe} simultaneously
drives each HDT toward its local mission objective,
rewards coordination links whose mutual information
exceeds their communication cost, and drops links whose
coordination benefit is negative. The network thereby
self-organizes its coordination topology. Links carrying
informative belief updates are maintained, links whose
mutual information falls below their communication cost
are released, and new links form when epistemic coupling
between agents exceeds the activation threshold. This
self-organizing property directly resolves the challenge of how network topology should be optimized
for epistemic rather than throughput objectives. In short, topology
is not optimized separately but emerges as a consequence
of collective free energy minimization.
\vspace{-3mm}\subsection{ Spatio-Temporal Integrated Information and Collective Intelligence}
Quantifying when collective intelligence genuinely emerges requires a measure that captures both the irreducibility of integration and the spatiotemporal structure of coordination. Inspired by IIT~\cite{tononi2016integrated}, which measures how much a system's collective behavior exceeds the sum of its parts, we define the \emph{spatiotemporal integrated information} for HDT-Net with joint sensing-communication-control  state $\mathbf{x} = \{(s_i, a_i, \mu_i)\}_{i=1}^{N}$ as:
\begin{equation}\small
\Phi(\mathbf{x}, \mathcal{R}, \Delta t) \;\!\!=\;\!\!\!\!\!\!
\min_{P \,\in\, \mathcal{P}_{\text{ST}}(\mathcal{R},\, \Delta t)}
\left[ \mathbb{I}(x^{\text{past}}; x^{\text{future}})
- \sum_{p \in P} \mathbb{I}(x_p^{\text{past}}; x_p^{\text{future}}) \right],
\label{eq:phi_st}
\end{equation}
where $\mathcal{P}_{\text{ST}}(\mathcal{R}, \Delta t)$ is the set of partitions that separate agents by spatial region~$\mathcal{R}$ and temporal synchronization window~$\Delta t$, with each partition
$P = \{p_1, p_2, \ldots, p_K\}$ dividing the $N$ HDTs
into $K$ disjoint groups, and
$\mathbf{x}_p = \{(s_i, a_i, \mu_i)\}_{i \in P}$
denotes the joint state of all HDTs assigned to
group~$p$. The mutual information terms $\mathbb{I}$ are computed over time-directed state transitions capturing causal dependencies rather than mere correlation. Specifically, the measure asks how much would collective inference degrade if agents within spatial proximity~$\mathcal{R}$ lost temporal synchronization beyond~$\Delta t$? When $\Phi_{\text{ST}}$ is high, the agents in that spatiotemporal region form a genuinely integrated cognitive unit whose coordination cannot be partitioned without significant inference loss. When $\Phi_{\text{ST}}$ is low, agents operate near-independently and coordination overhead outweighs its benefit, allowing the network to release communication resources for other uses. Critically, because the mutual information is computed over time-directed transitions, $\Phi_{\text{ST}}$ captures \emph{causal} dependencies and directed information flow, not mere correlation. For example, during disaster response wide-area search with spatially separated robots, low $\Phi_{\text{ST}} \approx 0$ indicates independent operation suffices since agents have uncorrelated observations within the region. During survivor extraction requiring multi-modal sensor fusion and human oversight, high $\Phi_{\text{ST}} \gg 0$ indicates tight integration is essential since any partition preventing coordination within the spatiotemporal window causes mission failure.

To reveal the structure of collective intelligence and guide sensing, communication, and computing resource allocation, $\Phi$ can be decomposed into three components:
$
\Phi \;=\;
\underbrace{\Phi_{s}}_{\substack{\text{integration across}\\ \text{co-temporal agents}}}
\;+\;
\underbrace{\Phi_{t}}_{\substack{\text{integration across}\\ \text{time within agents}}}
\;+\;
\underbrace{\Phi_{c}}_{\substack{\text{irreducible}\\ \text{spatiotemporal coupling}}}$.
$\Phi_{s}$ measures how much collective inference degrades when spatially separated agents are partitioned at a single time instant, capturing the value of fusing information from heterogeneous agents. $\Phi_{t}$ measures how much each agent's predictive capacity depends on its own historical trajectory, capturing the value of persistent world model memory. $\Phi_{c}$ is the component that cannot be attributed to either spatial or temporal integration alone. It quantifies the irreducible spatiotemporal coupling that arises when agents must not only observe complementary regions of the environment but do so in a temporally coordinated manner to maintain a coherent shared context. This cross term is the shared spatiotemporal context that is enabled HDT-Net. It vanishes when agents share spatial coverage but operate asynchronously, or when agents are temporally synchronized but observe redundant spatial regions.

\vspace{-2mm}\subsection{Intelligence Growth Through Coordinated
Learning}
When physical AI agents first enter a shared
environment, each HDT operates from its own local
observations with limited knowledge of neighboring
agents' states, yielding low $\Phi$
across the region. As agents begin exchanging beliefs
through wireless links, their world models
progressively align and complement one another,
resolving uncertainties that no single agent can
address alone. The network's objective is to
accelerate this transition from independent operation
to integrated collective cognition by maximizing the
rate at which $\Phi$ grows over the
mission horizon. The network's objective is to
accelerate this transition from independent operation
to integrated collective cognition by maximizing the
rate at which $\Phi_{\text{ST}}$ grows over the
mission horizon. This is captured by a growth-aware
objective that balances immediate physical AI task
performance against the development of collective
intelligence:
\vspace{-3mm}\begin{equation}
J = \underbrace{\sum_t r_t}_{\substack{\text{immediate}\\
\text{task reward}}}
+ \;\alpha\, \underbrace{\Phi(T)}_{\substack{
\text{terminal}\\\text{intelligence}}}
+ \;\beta \underbrace{\int_0^T
\dot{\Phi}(t)\,dt}_{\substack{
\text{intelligence}\\\text{growth}}}.
\label{eq:growth_objective}
\end{equation}
The first term drives each HDT toward its immediate
mission objective. The second term rewards the level
of spatiotemporal integration achieved by mission end,
ensuring the network does not sacrifice collective
coherence for short-term gains. The third term rewards
the rate of intelligence growth throughout the mission,
incentivizing early investment in coordination
structures that enable progressively more demanding
mission phases. The weights $\alpha$ and $\beta$
encode mission structure: when future coordination
failures are catastrophic and unrecoverable, $\beta$
is high and the network prioritizes intelligence
growth; when the mission is short and coordination
patterns are stable, immediate reward dominates.

The network drives $\Phi$ growth by
jointly optimizing across four resource dimensions.
Sensing resources are directed toward observations
that most reduce collective uncertainty about hidden
states within the spatiotemporal region, growing
$\Phi_{s}$ by filling gaps in spatial
coverage that no single agent can address.
Communication resources are optimized to prioritize belief exchanges
whose cognitive value is highest, growing
$\Phi_{c}$ by tightening temporal
synchronization among causally coupled agents.
Computation resources maintain and refine each HDT's
generative world model, growing $\Phi_{t}$
by extending the predictive horizon over which each
agent can anticipate future states. Control policies
coordinate agent trajectories and task assignments
to increase causal coupling where coordination is
needed, physically positioning agents to maximize
the spatiotemporal integration that sensing,
communication, and computation resources can then
sustain.
The central insight is that 6G physical AI networks are not static cognitive systems but evolving intelligences, and managing this evolution is as critical as optimizing instantaneous performance. Just as a network with high but static $\Phi$ cannot adapt when missions change or networks are disrupted, a network that continuously refines its collective world model through coordinated sensing, belief exchange, and causal discovery becomes increasingly capable over time, turning the network itself into a learning asset rather than a fixed infrastructure.

\section{Open Problems and Challenges of Networked Physical AI Systems}

\begin{itemize}

\item\emph{Challenge 1: Cognitive Traffic Generation and Belief-Centric Communication Metrics.}
In physical AI systems, traffic generation is driven by cognitive value, with agents transmitting only when communication is expected to reduce uncertainty of other agents world models or improve decision-making. As a result, traffic arrivals depend on evolving beliefs, uncertainty estimates, and coordination needs, rather than fixed sampling rates or exogenous processes. This leads to bursty, correlated traffic patterns, where transmission rates spike collectively when agents encounter novel scenarios requiring coordination. Characterizing such traffic requires new models in which transmission decisions optimize expected cognitive value in \eqref{eq:cog_value}, coupling arrival processes with distributed cognitive state evolution.
This necessitates a shift from traditional performance metrics such as bit error rate to \emph{belief fidelity}, which quantifies how effectively a received message reduces epistemic uncertainty at the receiving HDT and improves collective intelligence. Formally, this calls for a new rate--belief--distortion framework, where distortion is measured not by reconstruction error but by the KL divergence between the posterior achieved under the actual channel and that under an ideal noiseless channel. Such a formulation captures how wireless impairments degrade collective causal inference rather than signal fidelity. Moreover, it naturally induces a notion of \emph{cognitive capacity}, defined as the maximum achievable reduction in belief uncertainty per unit system resource, thereby jointly characterizing the impact of communication constraints, computational limitations, and inference dynamics on overall performance.

 \item \emph{Challenge 2: Latency degrades beliefs, and packet loss is a safety issue.} Classical networked control systems measure latency cost through age-of-information metrics  \cite{kosta2017age} quantifying how stale state estimates degrade control performance in linear dynamical systems. In contrast, physical AI agents must maintain internal world models
tightly synchronized to the real-world dynamics they
represent. When a belief update arrives late over a congested wireless channel, its value depends on how much the receiving agent's model has evolved during the delay interval due to its interactions with the world, not merely on the age of the underlying measurement.
During high-uncertainty periods when agents encounter novel scenarios, delayed beliefs may destabilize subsequent actions, potentially causing safety violations. During low-uncertainty periods when agents operate confidently within known environments, identical delays cause negligible degradation. Beyond
staleness, differential latency across multiple
communication links can cause agents to misinterpret
the temporal causality of events. When two correlated
belief updates arrive in an order that reverses their
true causal sequence, the receiving agent constructs
a world model in which effects appear to precede
their causes, producing control decisions grounded in
a physically impossible causal history \cite{popovski2024time}. For physical
AI systems operating under tight safety margins, such
causal misattribution means that the agent acts confidently on a world model
that is not merely stale but structurally incorrect. Existing age-of-information frameworks measure staleness uniformly and are unable to capture this coupling between network delay statistics and time-varying agent uncertainty. Here, an open problem is  defining world model belief age metrics that quantify information value decay as a function of both wireless channel delay and receiving agent's epistemic state evolution, then deriving bounds and scheduling policies that prioritize transmissions when delay costs are highest.
\item \emph{Challenge 3: Network topology determines collective intelligence.} In physical AI systems, the communication graph structure determines which agents share beliefs about which environment states. Fully connected topologies enable complete observation pooling, minimizing collective uncertainty but saturates wireless spectrum and energy budgets. Sparse topologies conserve resources but leave agents uninformed about environment regions they cannot directly sense. The optimal topology must minimize collective uncertainty subject to wireless channel capacity,  energy availability, and computational and sensing capabilities. They must adapt dynamically based on MB boundaries as mission requirements evolve and agents move through environments with time-varying channel conditions.
 While traditional methods minimize hop count, maximize throughput, or optimize coverage assuming fixed traffic demands, physical AI topology design must minimize an epistemic objective, specifically collective uncertainty over environment states, subject to wireless resource constraints across a combinatorial space of graph structures under mobility, fading, and interference. The optimization couples graph combinatorics with distributed probabilistic inference, where link addition or removal changes not just routing efficiency but the information fusion structure determining which agents can update which beliefs. No existing topology design framework addresses this \emph{coupling between network structure and collective epistemic state}.
\item \emph{Challenge 4: Network-Enabled Shared Temporal Context for Physically Coupled Swarms.} The collective gradient flow of \eqref{eq:collective_flow} describes networked HDTs evolving together as a single cognitive agent, but this description is only coherent if the network ensures that agents commit to coordinated actions within a synchronized spatiotemporal context window. Physical AI swarms executing tightly coupled tasks such as collaborative manipulation, formation flight, or synchronized assembly require the network to actively provide a shared temporal context in which all agents receive belief updates and commit to actions at the same physical instant. When this network-provided temporal alignment is absent, the swarm degenerates into a collection of individually rational agents whose actions are physically misaligned. The open problem is co-designing belief delivery, inference scheduling, and action commitment across the wireless infrastructure such that a network of HDTs acts as a single coordinated organism with deterministic timing guarantees, a requirement that none of the existing wireless synchronization standards address for cognitive multi-agent systems.
\item \emph{Challenge 5: Scalable Computation of
Spatiotemporal Integration.} Computing
$\Phi$ exactly requires evaluating mutual
information across all spatiotemporal partitions of
the HDT-Net, which is intractable for large
physical agent populations. Practical deployment of HDT-Net
requires scalable approximations that preserve the
measure's ability to distinguish genuine collective
intelligence from coordination overhead. Promising
directions include local approximations that compute
$\Phi$ over MB neighborhoods
rather than the full network, hierarchical
decomposition that exploits the holonic structure to
bound network-level integration from agent-level and
team-level estimates, and online incremental updates
that track $\Phi$ changes as agents enter
or leave spatiotemporal regions without full
recomputation. Initial evidence that IIT-based
measures are computable in practice comes from our
prior work on causal semantic
communication~\cite{ChristoJSAITArxiv2022}, where integrated
information was computed over structural causal models
to identify semantic content elements and quantify
irreducible integration, demonstrating feasibility
within individual agent neighborhoods. Extending
these results to network-scale $\Phi$
across hundreds of agents remains an open challenge.  The key constraint is that the
approximation error must remain small enough to
preserve correct resource allocation decisions,
since overestimating $\Phi$ wastes
communication resources on unnecessary coordination
while underestimating it causes agents to operate
independently when collective inference is essential
for safe physical AI operation.
\item \emph{Challenge 6: Adversarial Resilience and Trust
in Holonic Coordination.} When one or more HDTs are
compromised by adversarial attack, the beliefs they
transmit across MB boundaries may be
intentionally misleading, corrupting neighboring
agents' world models and degrading collective
intelligence. While the blanket structure confines
direct influence to causal neighbors, a compromised
agent at a critical position in the coordination
topology can propagate corrupted beliefs through
chains of legitimate updates. Open problems include a)
developing distributed trust mechanisms where each
HDT evaluates the consistency of received beliefs
against its own local observations and the
predictions of its theory of mind functor, b) designing
blanket reconfiguration protocols that isolate
suspected adversarial agents by dynamically
restructuring coordination boundaries, and
c) quantifying the resilience of $\Phi$ under
partial compromise to determine how many agents can
be corrupted before collective intelligence degrades
below mission-critical thresholds. Unlike
conventional network security that protects data
integrity at the bit level, adversarial resilience
in HDT-Net must protect \emph{belief integrity} at the
semantic level, ensuring that the collective world
model remains causally consistent even when
individual agents inject false causal claims about
the physical environment.
\end{itemize}
\vspace{-1mm}\section{Conclusion}
\label{sec:conclusion}

In this paper, we have introduced   HDT-Net, a framework that
transforms wireless networks from passive data
transport into holonic reasoning systems for
physical AI, where each twin operates as an
autonomous cognitive agent while composing into
progressively larger collective units that adapt
and grow smarter than any individual agent can
alone. By grounding network partitioning in
causal MBs, ensuring compositional
consistency through category theory, unifying
perception-action-learning through active inference,
and quantifying collective intelligence through
spatiotemporal integrated information
$\Phi_{\text{ST}}$, HDT-Net enables networks that
reason about what they do not know, transmit beliefs
whose cognitive value is measured by their impact on
the receiver's next physical action, and actively
cultivate collective intelligence as a first-class
network resource. The transition from 5G to 6G is not
about faster pipes but about networks that understand
uncertainty, coordinate through meaning rather than
bits, and grow collectively smarter over time. HDT-Net
provides the theoretical foundation for this
transformation.

\bibliographystyle{IEEEtran}
\def\baselinestretch{0.84}\bibliography{references}

\end{document}